\documentclass[11pt,a4paper]{article}
\usepackage[left=15mm, top=10mm, right=15mm, bottom=20mm, nohead=true, nofoot=true]{geometry}

\usepackage[utf8]{inputenc}
\usepackage{authblk}
\usepackage{indentfirst}
\usepackage{misccorr}
\usepackage{graphicx}
\usepackage{amsmath}
\usepackage{amssymb}
\usepackage{multicol}
\usepackage{bm}
\usepackage{longtable}
\usepackage{pdflscape}
\usepackage{epstopdf}
\usepackage{multirow}
\usepackage{adjustbox}
\usepackage{amsfonts}
\usepackage{ifthen}
\usepackage{fancyhdr}
\usepackage{setspace}
\usepackage{xcolor}
\usepackage[round,authoryear,comma,sort&compress,numbers]{natbib}
\usepackage[toc,title,page]{appendix}
\usepackage[hidelinks]{hyperref}
\usepackage{url}

\hypersetup{
    colorlinks=true,
    linkcolor=green,
    citecolor=blue,
    filecolor=magenta,
    urlcolor=cyan,
    pdftitle={Sharelatex Example},
    pdfpagemode=FullScreen,
}

\fancypagestyle{plain}{\chead{\texttt{\textcolor{brown}{Communications of BAO, Vol. 72, Issue 2, 2025, pp. 351-357}}}}
\title{\textbf{Systematic and Statistical Uncertainties in Cepheid PL Relations: Incorporating a Cross-Filter Random-Phase Mitigation Approach}}

\author[1]{M. Abdollahi \thanks{m.abdollahi@ipm.ir, Corresponding author}}
\author[1]{A. Javadi \thanks{atefeh@ipm.ir}}

\affil[1]{\scriptsize School of Astronomy, Institute for Research in Fundamental Sciences (IPM), Tehran, 19568-36613, Iran}

\begin{document}
\pagestyle{empty}
\newpage
\pagestyle{fancy}
\label{firstpage}
\date{}
\maketitle
\begin{abstract}
The Period–Luminosity (PL) relation of Cepheid variable stars is a fundamental tool for measuring extragalactic distances and constraining the Hubble constant ($H_0$). Achieving high precision in PL-based distances requires careful consideration of both systematic and statistical uncertainties. We review the main sources of these uncertainties in PL relations, highlighting the increasing impact of random-phase errors in single-epoch observations from limited temporal coverage, such as those obtained with the James Webb Space Telescope (JWST). We discuss mitigation strategies for systematic errors, including photometric calibration offsets, metallicity effects, blending, and parallax biases, and quantify key contributors to statistical errors, such as photometric noise, intrinsic scatter, and phase-sampling limitations. Special attention is given to a recently proposed cross-filter random-phase correction method \citep{Abdollahi_2025_RPC}, which recovers mean magnitudes from single-epoch data by exploiting correlations between PL residuals in different bands. This technique reduces the dispersion in the infrared PL relation by 28\%, equivalent to an order-of-magnitude increase in effective temporal sampling, demonstrating an efficient path to improving Cepheid-based distance measurements and the precision of $H_0$.
\end{abstract}


\section{Introduction}

Henrietta Leavitt’s law, or the Period–Luminosity (PL) relation of Cepheid variable stars, has been known for over a century and is a fundamental rung in the cosmic distance ladder\citep{Leavitt_1912_Cep_PLR}. It enables precise distance measurements to nearby galaxies, which in turn calibrate secondary distance indicators such as Type Ia supernovae, ultimately contributing to the determination of the Hubble constant ($H_0$) \citep{Freedman_2023_Progress,Riess_2024_IAUS}. Even small uncertainties in the zero point or slope of the PL relation propagate directly into $H_0$, highlighting the need for a thorough understanding of the uncertainties in PL analyses \citep{Freedman_2012_MidIR}.

Errors in the Cepheid PL relation can be broadly categorized as systematic or statistical (random) \citep{Owens_2022_Challenges,Riess_2016_2.4H0}. Systematic errors affect multiple stars in a correlated manner and include photometric zero-point offsets, calibration differences between instruments, metallicity effects, residual extinction, crowding in dense stellar fields, and parallax zero-point biases \citep{Madore_2024_Systematics,Freedman_2012_MidIR}. Statistical errors vary independently from star to star and generally decrease with increasing sample size \citep{Madore_2009_Second_Epoch,Abdollahi_2025_RPC,Madore_2024_Systematics}. Their main contributors are photometric noise, phase-sampling limitations, and the intrinsic scatter in Cepheid luminosities.

Different photometric bands are subject to varying levels of uncertainty due to factors such as extinction, metallicity, and crowding. For example, observations in optical and near-optical filters are particularly sensitive to these effects, which can introduce significant systematic uncertainties in the derived magnitudes and the calibration of the PL relation \citep{Tanvir_1997_Cepheids}. In contrast, infrared observations mitigate many of these issues: they are less affected by interstellar extinction, exhibit smaller amplitude variations, and show a tighter intrinsic PL relation, thereby improving the overall precision of distance estimates \citep{Freedman_2001_Final_HST,Persson_2004_nearIR}. Nonetheless, both systematic and statistical errors remain non-negligible in the infrared domain and must be carefully quantified to ensure reliable PL-based distances \citep{Freedman_2012_MidIR,Freedman_2010_H0}.

This study examines uncertainties in the PL relations, with a particular focus on random-phase statistical errors. Section \ref{sec:sys} discusses the main sources of systematic uncertainty and strategies to mitigate them, while Section \ref{sec:statistic} highlights statistical errors, emphasizing random-phase errors as a significant contributor. Section \ref{sec:mitigating} reviews recent work by \cite{Abdollahi_2025_RPC}, which demonstrates a correction method for random-phase errors. Finally, Section \ref{sec:conclusion} summarizes the main findings and implications for high-precision distance measurements.

\section{Systematic Errors}\label{sec:sys}

Systematic errors in the Cepheid PL relation are coherent shifts in the measured magnitudes or distances of multiple stars. Unlike statistical uncertainties, they do not decrease with increasing sample size and therefore set the fundamental limit on precision. Even in the infrared, where extinction and intrinsic scatter are minimized, systematics remain the dominant source of uncertainty.

The main contributors to systematic errors and corresponding mitigation strategies are summarized below:

\begin{itemize}

\item \textbf{Photometric Calibration Offsets:} 
Systematic offsets in measured magnitudes can arise from differences in photometric calibration between instruments or datasets. This effect can be expressed as

\begin{equation}
\Delta m_{\rm sys} = ZP_{\rm inst} - ZP_{\rm std},
\end{equation}

where $ZP_{\rm inst}$ and $ZP_{\rm std}$ are the instrumental and standard photometric zero points, respectively. Such offsets represent systematic differences in the measured magnitudes due solely to calibration, independent of the intrinsic properties of the stars or the PL relation. To mitigate these systematic shifts, multiple instruments can be cross-calibrated using standard stars, consistent photometric systems can be adopted, and zero points can be verified through overlapping observations \citep{Regnault_2009_Photometric_calibration}.

\item \textbf{Metallicity Dependence:} 
Cepheid luminosities at a given period are slightly affected by metallicity:
\begin{equation}
M = a \log P + b + \gamma \left({\rm [Fe/H]} - {\rm [Fe/H]}_0\right),
\end{equation}
where $\gamma$ quantifies the sensitivity to metallicity.  
To mitigate this effect, one may apply empirical or theoretical metallicity corrections, or alternatively focus on infrared bands, which are less sensitive to metallicity, although the effectiveness of this latter approach remains debated \citep{Breuval_2022_Metallicity,Trentin_2024_CMetall}.

\item \textbf{Residual Extinction and Reddening Law Uncertainties:} 
Even after standard extinction corrections, imperfect knowledge of the reddening law can introduce systematic offsets in Cepheid magnitudes. One way to reduce this effect is through the use of Wesenheit magnitudes \citep{madore_1982_Wesenheit}, defined as

\begin{equation}
W = m_\lambda - R_{\lambda_1, \lambda_2} \left(m_{\lambda_1} - m_{\lambda_2}\right),
\end{equation}

where $R_{\lambda_1, \lambda_2}$ is the total-to-selective extinction ratio. This systematic uncertainty can be further mitigated by modeling the reddening law for each galaxy individually or by applying the PL relation at longer wavelengths, particularly in the infrared \citep{Freedman_2001_Final_HST}.

\item \textbf{Blending and Crowding:} 
In dense stellar fields, unresolved companions or background stars artificially brighten measured Cepheid magnitudes:
\begin{equation}
F_{\rm obs} = F_{\rm Cepheid} + F_{\rm blend} \quad \Rightarrow \quad 
m_{\rm obs} = -2.5 \log(F_{\rm obs}).
\end{equation}
This systematic can be minimized by employing high-resolution imaging (e.g., \textit{HST} or \textit{JWST}) or performing artificial-star tests to quantify blending. Statistical deblending techniques can also be applied in wide-field surveys to correct measured magnitudes \citep{Riess_2024_JWST_Crowding}.

\item \textbf{Distance Calibration and Parallax Systematics:} 
Systematic errors in parallax-based distances, such as zero-point offsets in \textit{Gaia} or \textit{HST}, propagate into the PL zero point and slope \citep{Casertano_2017_Parallax}.  
These uncertainties can be minimized by using high-precision parallax measurements, such as Gaia DR4 (expected in 2026), applying the corresponding zero-point corrections, and combining independent distance calibration methods, such as eclipsing binaries or maser distances.

\end{itemize}

By carefully identifying and addressing each of these systematic effects, the total uncertainty in Cepheid-based distance measurements can be significantly reduced, enabling more precise determinations of the Hubble constant ($H_0$) and other cosmological parameters.

\section{Statistical Errors} \label{sec:statistic}

Statistical errors in the Cepheid PL relation represent random uncertainties that vary from star to star and, unlike systematic effects, can be reduced by increasing the sample size. In the infrared, where extinction and intrinsic scatter are minimized, the main statistical sources of uncertainty include photometric measurement noise, intrinsic luminosity dispersion, residual random extinction, and incomplete phase coverage.

The total per-star statistical uncertainty can be expressed as the quadrature sum of these components:
\begin{equation}
\sigma_{\rm star} = \sqrt{\sigma_{\rm phot}^2 + \sigma_{\rm phase}^2 + \sigma_{\rm intr}^2 + \sigma_{\rm ext,rand}^2},
\end{equation}
where $\sigma_{\rm phot}$ is the photometric error, $\sigma_{\rm phase}$ accounts for incomplete light-curve sampling, $\sigma_{\rm intr}$ is the intrinsic scatter, and $\sigma_{\rm ext,rand}$ represents random extinction.  

For a sample of $N$ independent Cepheids, the statistical uncertainty in the PL zero point can be expressed as \citep{Taylor_1997_error_analysis,Bevington_2003_error_analysis}  

\begin{equation}
\sigma_{\rm ZP} = \frac{\overline{\sigma_{\rm star}}}{\sqrt{N}},
\end{equation}

where $\overline{\sigma_{\rm star}}$ is the mean per-star uncertainty. This uncertainty propagates into the distance modulus and fractional distance error according to
\begin{equation}
\delta_d \approx 10^{\sigma_{\rm ZP}/5} - 1 \approx 0.4605 \, \sigma_{\rm ZP} \quad (\text{for small } \sigma_{\rm ZP}).
\end{equation}

The main sources of statistical uncertainty and their mitigation strategies are outlined below.

\begin{itemize}

\item \textbf{Photometric Measurement Noise:}  
Random noise in the measured magnitudes broadens the PL relation and directly contributes to $\sigma_{\rm phot}$. For a single flux measurement $F$ with uncertainty $\sigma_F$, the corresponding photometric error can be expressed as \citep{Howell_2000_Handbook}:
\begin{equation}
\sigma_{\rm phot} = \frac{2.5}{\ln 10} \frac{\sigma_F}{F}.
\end{equation}
This term can be minimized by obtaining high signal-to-noise observations, improving background subtraction, and performing accurate detector calibration.

\item \textbf{Intrinsic Luminosity Scatter:}  
At a fixed period and metallicity, Cepheids exhibit an intrinsic dispersion in luminosity due to physical differences in temperature, mass, and pulsation mode:
\begin{equation}
M_\lambda = a \log P + b + \epsilon_{\rm intr},
\end{equation}
where $\epsilon_{\rm intr}$ represents the intrinsic scatter with variance $\sigma_{\rm intr}^2$. Observations in the infrared mitigate this effect because luminosity variations from temperature changes are smaller at longer wavelengths.

\item \textbf{Residual Random Extinction:}
Spatially varying dust introduces random star-to-star magnitude fluctuations, which can be expressed as

\begin{equation}
\sigma_{\rm ext,rand} \approx R_\lambda \, \sigma_{E(B-V)},
\end{equation}

where $R_\lambda$ is the total-to-selective extinction ratio, and $\sigma_{E(B-V)}$ represents the local dispersion in color excess. This source of uncertainty can be minimized through multi-band photometry and the application of local reddening corrections \citep{Nataf_2013_Dust}.

\item \textbf{Incomplete Phase Coverage:}  
Sparse or irregular temporal sampling introduces phase-dependent magnitude errors, leading to increased scatter in the PL relation \citep{Yuan_2022_JWST}. For a single-epoch observation, the uncertainty on the intensity-mean magnitude can be approximated as
\begin{equation}
\sigma_{\rm phase} \approx A_\lambda \, f(\phi),
\end{equation}
where $A_\lambda$ is the light-curve amplitude in band $\lambda$ and $f(\phi)$ is a phase-dependent correction factor. The effect can be reduced by obtaining well-sampled light curves or applying techniques to reconstruct the mean magnitude from limited data.

\end{itemize}

Infrared observations offer the additional advantage of simultaneously reducing intrinsic scatter and random extinction, allowing total statistical uncertainties in the infrared PL relation to approach sub-percent levels. Achieving such precision, however, remains observationally demanding. Limited phase coverage, especially in space-based infrared programs such as those conducted with the \textit{JWST}, often increases $\sigma_{\rm phase}$ and thus the total error budget \citep{Gardner_2023_JWST}. In the following section, we examine in detail the impact of incomplete phase coverage, or random-phase uncertainty, and discuss methods proposed to mitigate its effects.

\subsection{Random Phase Errors}

Limited phase sampling of Cepheids causes the measured magnitudes to deviate from their true mean values, introducing additional scatter into the PL relation. This effect is particularly relevant for space-based infrared observations, where limited temporal coverage often prevents complete phase sampling. The random phase error for a single star can be approximated as:
\begin{equation}
\sigma_{\rm phase} = \frac{A_\lambda }{\sqrt{N_{\rm obs}}},
\end{equation}
where $A_\lambda$  is the amplitude of the light curve in the observed band, and $N_{\rm obs}$ is the number of independent observations sampling the light curve. This relation reflects the fact that more observations distributed over the pulsation period reduce the uncertainty on the mean magnitude.

To mitigate random-phase errors, several strategies can be employed, some of which have been demonstrated in previous studies:

\begin{itemize}
\item \textbf{Dense temporal sampling:} Observing Cepheids at multiple, evenly distributed epochs over their pulsation periods minimizes phase-related scatter. This approach is straightforward and represents the most direct method for reducing such errors.

\item \textbf{Phase-weighted averaging:} Assigning weights to individual measurements based on phase coverage reduces bias in the estimation of mean magnitudes \citep{Saha_1990_weighted_phase}.  

\item \textbf{Infrared observations:} Since infrared light curves exhibit smaller amplitudes than their optical counterparts, the phase-induced dispersion ($\sigma_{\rm phase}$) is naturally reduced, making infrared bands particularly advantageous for sparsely sampled datasets \citep{Freedman_2012_MidIR}.  

\item \textbf{Template fitting:} Employing well-calibrated light-curve templates enables the reconstruction of intensity-mean magnitudes even from sparsely sampled data \citep{Soszy_2005_Mean_Single_epoch, Riess_2023_Single_epoch_JWST}.

\end{itemize}

By properly accounting for random phase errors and adopting these mitigation techniques, the contribution of phase uncertainty to the total statistical error in the PL zero point can be significantly minimized, improving the precision of Cepheid-based distance measurements.

In addition to the approaches discussed above, two methods have been proposed to correct for the effects of random-phase sampling \citep{Abdollahi_2025_RPC,Madore_2024_NovelMethod}, the first of which is summarized in the following section.

\begin{figure}
        \vspace{0.5cm}
        \centering
        \includegraphics[width=0.6\linewidth]{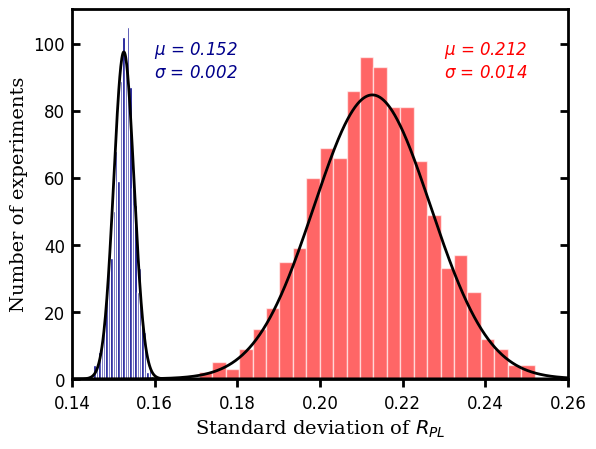}
        \caption{Adapted from \cite{Abdollahi_2025_RPC}. Distribution of residuals from the PL relation before (red) and after (blue) applying the random-phase correction. Both distributions are approximately Gaussian, and the correction reduces the standard deviation from $0.212$ mag to $0.152$ mag, demonstrating a significant improvement in the precision of single-epoch J band magnitudes.}
        \label{fig:Std_PLR}
\end{figure} 

\section{Mitigating Random-Phase Sampling Noise in the Cepheid Period-Luminosity Relation: A Cross-Filter Consistency Approach} \label{sec:mitigating}

\cite{Abdollahi_2025_RPC} proposed a method to mitigate the uncertainties in single-epoch observations arising from random-phase sampling. The approach exploits two intrinsic correlations: (i) a direct correlation between the residuals of the PL relations in two photometric bands, and (ii) a cross-correlation between the PL residuals and the residuals from the magnitude-magnitude (MM) diagram constructed from the same pair of bands. The correction is formulated as

\begin{equation} \label{eq:correction_raw}
\begin{split}
cm(i',j) = m(i',j) &- \alpha_{0} \times R_{\mathrm{PL}}(k,j) - \alpha_{1} \times R_{\mathrm{MM}}(i'k,j) - \beta_{0},
\end{split}
\end{equation}

where $m(i',j)$ is the observed single-epoch magnitude of star $j$ in random-phased band $i'$, $R_{\mathrm{PL}}(k,j)$ denotes the residual of the PL relation in the reference band $k$, and $R_{\mathrm{MM}}(i'k,j)$ represents the residual in the MM relation between bands $i'$ and $k$. The coefficients $\alpha_{0}$, $\alpha_{1}$, and $\beta_{0}$ are determined empirically through extensive Monte Carlo simulations designed to minimize the dispersion of the corrected magnitudes.

Given the reduced intrinsic scatter and smaller light-curve amplitudes of infrared PL relations, the method was validated using the $J$ band as the primary filter and the $B$ band as the secondary one. After more than one thousand simulated realizations, the optimal coefficients and intercept were derived as follows:

\begin{equation} \label{eq:correction_final}
\begin{split}
cm(J',j) &= m(J',j) - (0.525 \pm 0.035), R_{\mathrm{PL}}(B,j) - (0.643 \pm 0.039), R_{\mathrm{MM}}(J'B,j) - (0.000 \pm 0.006).
\end{split}
\end{equation}

The application of the correction to single-epoch $J$ band magnitudes substantially tightened the PL relation, decreasing the dispersion from 0.212 mag (random-phase) to 0.152 mag (see Figure \ref{fig:Std_PLR}). This corresponds to a $\sim$28\% reduction in the statistical uncertainty of the apparent distance modulus.

The analysis further showed that the total variance in the random-phase PL relation arises equally from the intrinsic scatter of the time-averaged relation and the additional noise due to random sampling. Achieving a similar improvement by increasing the number of observations would require roughly an order of magnitude more epochs, demonstrating that this correction offers an efficient alternative for improving precision with minimal observational effort.

\section{Conclusion and Summary}\label{sec:conclusion}

The Period–Luminosity (PL) relation of Cepheid variable stars remains a cornerstone in the determination of extragalactic distances and the calibration of the Hubble constant. Achieving high precision in such measurements requires a comprehensive understanding and control of both systematic and statistical uncertainties. In this work, we reviewed the dominant sources of these uncertainties in the PL relations of Cepheids and discussed several methods to mitigate their effects.

Systematic errors, arising from calibration offsets, blending, metallicity effects, extinction corrections, and parallax biases,represent correlated uncertainties that cannot be reduced by increasing the sample size. Their mitigation requires consistent photometric calibration, accurate metallicity and reddening corrections, and the use of well-characterized parallax data such as Gaia DR4. Statistical errors, on the other hand, result from random effects such as photometric noise, intrinsic scatter, and incomplete phase coverage. Among these, random-phase sampling has emerged as a major contributor to the observed dispersion in single-epoch infrared observations, particularly those obtained with limited temporal coverage from space-based facilities like \textit{JWST}.

To address this challenge, a cross-filter random-phase correction method was recently introduced \citep{Abdollahi_2025_RPC}, exploiting correlations between PL residuals across different bands. This approach effectively reduces the dispersion in the PL relation without requiring extensive phase coverage, achieving an improvement equivalent to increasing the number of observations by an order of magnitude. Such corrections thus provide a powerful means of enhancing the precision of single-epoch photometry and improving the reliability of infrared PL-based distance determinations.

Continued refinement of these methods, combined with forthcoming high-precision data from missions such as \textit{JWST} and future infrared observatories, will further tighten the Cepheid PL relation. We hope that progress in fully identifying and accounting for all forms of errors associated with the Cepheid distance scale will advance our broader confidence in, and understanding of, the cosmic distance scale.

\section*{Acknowledgment}
We express our sincere gratitude to Barry F. Madore and Wendy L. Freedman for their valuable guidance, constructive comments, and supportive discussions during the preparation of this review. Their insights significantly improved the clarity and depth of this work.

\scriptsize
\bibliographystyle{ComBAO}
\nocite{*}
\bibliography{references}

\end{document}